\documentclass[12pt, draftclsnofoot, onecolumn]{IEEEtran}
\usepackage{setspace}
\usepackage{amsmath,amssymb,amscd,latexsym,dsfont}
\usepackage{float,color,graphicx,subfigure}
\usepackage{changes}
\usepackage{comment,cite}
\usepackage{enumerate}
\usepackage{stfloats}
\usepackage{psfrag}
\usepackage{cite}
\usepackage{epstopdf}
\usepackage{booktabs}
\usepackage{fleqn}
\usepackage{multirow}

\newcommand{\mb}[1]{\boldsymbol{#1}}

\definecolor{mygreen}{rgb}{0.0, 0.56, 0.0}

\newtheorem{example}{Example}
\newcommand{\thickhline}{\noalign{\hrule height 1.1pt}}
\begin{document}
\sloppy
\title{Efficient Multi-Dimensional Mapping Using QAM Constellations for BICM-ID}

\author{
\IEEEauthorblockN{Hassan M. Navazi  and Md. Jahangir Hossain, \emph{Member, IEEE}\\}
\IEEEauthorblockA{
The University of British Columbia, Kelowna, BC, Canada\\
\emph{hnavazi@alumni.ubc.ca},   \emph{jahangir.hossain@ubc.ca}
}
}
\maketitle
\begin{abstract}
\textbf{Bit-interleaved coded modulation with iterative decoding (BICM-ID)  offers  very good error performance over   additive white Gaussian noise (AWGN) and fading channels if it uses a wisely designed signal mapping. Further,  error performance of BICM-ID can  significantly be improved by employing multi-dimensional (MD) modulation. However,  suitable  MD mappings   are obtained by computer search techniques   except for  MD modulations that use smaller constellation   e.g.,  binary phase shift keying (BPSK), quadrature phase shift keying (QPSK) and $8$-ary phase shift keying ($8$-PSK) as basic modulation. The alphabet size of MD modulations  increases exponentially as the order of the basic modulation increases and computer search  becomes intractable. 
In this paper, we propose a systematic  mapping method for MD modulations. The innovativeness of our proposed method is that  it    generates  MD mappings  using $16$- and $64$-quadrature amplitude modulation (QAM)  very efficiently. The presented numerical results show that the proposed method  improves  bit error rate (BER) of BICM-ID.}
\end{abstract}


\section{Introduction}
\label{sec:intro}
In \cite{Zehavi}, Zehavi proposed bit-interleaved coded modulation (BICM) which was analytically investigated by G. Caire \emph{et al} \cite{BICM}. BICM uses a bit-interleaver to separate the modulator from the encoder. Due to this separation, the modulator can be chosen independently from the encoder and it increases  design flexibility \cite{BICM}. It also improves the time diversity order of coded modulation \cite{Zehavi}.  Although random interleaving improves performance of BICM over fading channels, it reduces the  minimum  Euclidean distance between signals and degrades performance over additive white Gaussian noise (AWGN) channels \cite{8PSK_signaling}.  To overcome this problem, iterative decoding was proposed for BICM receiver \cite{BICMID}-\cite{Benedetto}. The resulted system is known as BICM with iterative decoding (BICM-ID) and achieves a significant coding gain through  iterations. As such a good performance is obtained over both AWGN and fading channels \cite{BICM-ID-good-AW-Ray}. 

 Symbol mapping is defined as labeling of constellation symbols with binary digits. It is well-known that \color{black}the performance of BICM-ID  depends on the applied symbol mapping  and a number of works has been carried out to address this issue, see for examples, \cite{BICM-ID-good-AW-Ray}-\cite{MDQAM}. 
   If a sequence of bits is mapped to a vector of symbols (symbol-vector)  rather than a single symbol, 
 the mapping is referred to as 
  multi-dimensional (MD) mapping \cite{MD-BQPSK-Simoen}.
  For example, if $N$ symbols from  phase shift keying (PSK) or quadrature amplitude modulation (QAM) are used in 
 each symbol-vector, a $2N$-D modulation is obtained. In fact, error performance of BICM-ID can be
significantly improved by employing MD modulation due to the increased 
  Euclidean distance between symbol-vectors \cite{MD-BQPSK-Simoen}, \cite{MDQAM}.\color{black}
   
Although a higher order MD modulation  offers more flexibility in generating good mappings for BICM-ID  \cite{MD-Hyper-Ha}, it tremendously increases the number of possible mappings. In particular,  $2N$-D modulations using $2^m$-QAM  symbols have  $2^{mN}!$  possible mappings where $!$ denotes a factorial operation.      So,  it is difficult  to find good/optimum mappings for   MD modulations. This problem is  more severe when  MD modulations are constructed using  higher order 2-D modulations. For example,  the number of possible mappings for $4$-D modulation using $16$-QAM  approaches infinity. 
  Suitable  MD mappings   are obtained by computer search techniques except for  MD modulations that use smaller constellation   e.g.,  binary PSK (BPSK), quadrature PSK (QPSK) and $8$-PSK as basic modulation.
 The so-called binary switch algorithm (BSA) \cite{BSA} is  the best known computer search method for finding good mappings.  However to obtain  suitable mappings for larger modulations such as MD modulations, the BSA becomes  intractable due to its  complexity \cite{rndm_map1}. In \cite{rndm_map1} and \cite{rndm_map2}, authors demonstrated  that random mapping can lead  to  efficient MD mappings. According to  the random mapping  technique,   computer search is used to obtain  a good mapping from a large  set  of randomly generated mappings that makes the procedure complex. Moreover, it degrades the resulted mappings' performance  especially for larger MD modulations.    

In this paper, we propose an efficient mapping method for MD modulations that use $2^{m}$-QAM ($m=4,6$) as basic modulation.   Our goal is to obtain mappings which achieve a lower error rate for BICM-ID at low SNR values as well as at high SNR values. A similar objective  is considered in \cite{MSED} where the authors used  a doping technique (combining two mappings) to obtain mappings for  $2$-D modulations. However,  instead of combining mappings,  we develop a single mapping for a given MD modulation  to  achieve a lower error rate for BICM-ID at low SNR values as well as at high SNR values.  Furthermore, our proposed method yields  mappings for MD constellations rather than $2$-D ones.   The proposed method is a heuristic-based technique and does not employ any computer search. The presented numerical results show that our approach \color{black}  not only yields mapping efficiently but also improves bit error rate (BER) performance of BICM-ID over both AWGN and Rayleigh fading channels. For example, it can save about $3$ dB transmit signal energy for a target BER of $10^{-6}$ compared to   the mappings found by the BSA and random mappings.\color{black}

The rest of the paper is organized as follows. While in Section \ref{sysmod} we provide BICM-ID system model, in Section \ref{Guide-Lines} we describe the design criteria of  MD mappings for BICM-ID  over AWGN and Rayleigh fading channels. In Section \ref{proposed_map}, we describe our  proposed MD mapping method.  In Section \ref{num_result}, we  present numerical results and
compare the  performance of our resulted  mapping.  Finally, Section \ref{conc.} concludes the paper. \\
\emph{Notations:} Throughout the paper, we use a lowercase letter e.g.,  $x$ to represent a variable, a boldface letter e.g.,  $\mb{x}$ to denote  a row vector,  and $x^{(i)}$ to represent  $i$th element of $\mb{x}$. The norm of a vector is denoted by $\Vert \mb{x}\Vert$ which is  defined  as $\Vert \mb{x} \Vert^{2} = \sum_{i=1}^{N} x_{i}^{2}$. Blackboard bold letters $\mathbb{E}$ and $\mathbb{O}$ denote the set of even integers and  the set of odd integers, respectively.   All-ones vector is also represented by $\mb{1}$. \color{black}
\color{black}
\section{System Model}
\label{sysmod}
\begin{figure}[t]
\begin{center}
\includegraphics[width=.8\columnwidth,viewport= 45mm 206mm 155mm 254.00mm,clip]{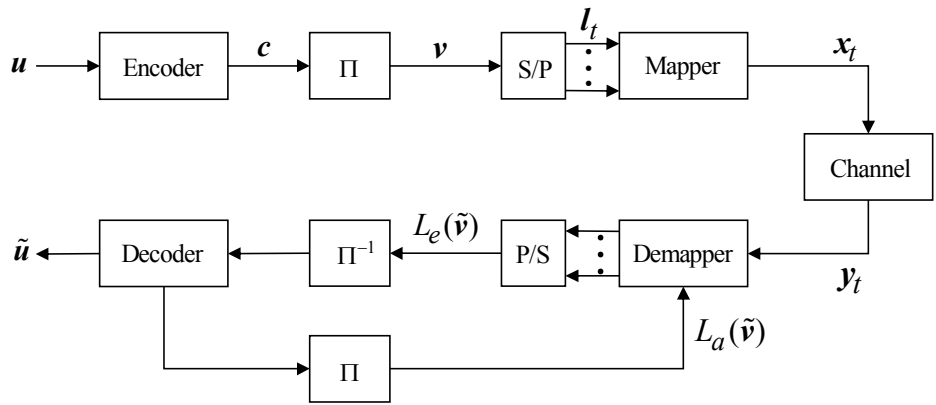} \caption{ The block diagram of a BICM-ID system.} 
\label{sys_mod}
\end{center}
\end{figure}

 A BICM-ID  system model is shown in Fig. \ref{sys_mod} where 
 a sequence of information bits ${\mb u}$ is encoded by a convolutional encoder. Then, the coded bits ${\mb c}$ are  randomly interleaved and the interleaved coded bits ${\mb v}$ \color{black}are grouped in blocks of $mN$ bits. For notational convenience, let us denote $t$th block  of interleaved coded   bits at the input of the modulator by $\mb{l}_{t}=[l_{t}^{(1)}, l_{t}^{(2)},\cdots, l_{t}^{(mN)}]$. The modulator maps ${\mb l}_{t}$ to a vector of $N$ consecutive $2^{m}$-ary signals, ${\mb x}_{t} = [x_{t}^{(1)}, x_{t}^{(2)}, \cdots, x_{t}^{(N)}]$, using a MD mapping function $\mu : \lbrace0, 1\rbrace^{mN}\longrightarrow \mb{\chi} = \chi^{N}$ where $\chi$ denotes the $2$-D $2^{m}$-ary signal set.  Mathematically,  we can write 
\begin{equation}
{\mb x}_{t} = [x_{t}^{(1)}, x_{t}^{(2)}, \cdots, x_{t}^{(N)}] =\mu({\mb l}_{t}) .
\end{equation}

 The average energy per symbol-vector is assumed to be $1$, i.e., $E_{\mb{x}_{t}}=1$.  
  At the receiver, the received signal vector corresponding to the transmitted symbol-vector  $\mb{x}_t$ can be expressed as 

\begin{equation}
{\mb y}_t = \mb{h}^{T}_t{\mb x}_t + {\mb n}_t,
\end{equation}where ${\mb h}_t = [h_{t}^{(1)}, h_{t}^{(2)}, \cdots, h_{t}^{(N)}]$ is the corresponding vector of the Rayleigh fading coefficients, $A^{T}$ represents the transpose of $A$, \color{black}  and ${\mb n}_t$ is a vector of $N$ additive complex white Gaussian noise samples with zero-mean and variance $N_0$. We consider two types  of fading channels as follows: (i) all $N$ symbols in a symbol-vector experience a constant fading coefficient, i.e.,   block-fading   channel \cite{Block-Fading1}-\cite{Block-Fading2} \color{black}and  (ii) the coefficients vary over the consecutive symbols in a symbol-vector, i.e., fast fading channel.  Clearly, $\mb{h}_t=\mb{1}$ corresponds to the AWGN channel. For brevity, we omit the index $t$ in subsequent  sections.   It is assumed that the receiver
has the perfect channel state information (CSI).

At the receiver, the demapper uses the received signal $\mb{y}_{t}$ and the \textit{a priori} log-likelihood ratio (LLR) of the coded bits to compute the extrinsic LLR for each of the bits in the received symbol as described in \cite{Brink}. Next, the random deinterleaver permutes the extrinsic LLRs which are used by  the channel decoder. The decoder  uses the BCJR algorithm \cite{BCJR} to calculate the extrinsic LLR values of  the coded bits. These LLR values are interleaved and fed back to the demapper that uses  them as the \textit{a priori} LLR values in the next iteration. \color{black}

\section{Motivation and Design Criteria of MD mappings for BICM-ID}
\label{Guide-Lines}

In low SNR region,   the performance of BICM-ID  depends on the    
BER  at the  first iteration, i.e.,  BICM-ID is equivalent to BICM.  So,  the optimum mapping for BICM-ID in low SNR region corresponds to  the optimum mapping for BICM. 
However, the optimal mapping designed for  low SNR region   is  not suitable for BICM-ID in  high SNR region. The reason can be explained as follows.  Although the optimal mapping designed for low SNR region provides an early turbo cliff in BER performance, it offers a poor error-floor which reaches the BER range of  practical interest, e.g., $10^{-6}$ at a very high SNR value.
 On the other hand,  the  mapping, which  is designed to minimize  the asymptotic error rate  of  BICM-ID, results in  extremely  low BER  at a very high SNR value.  Hence,  the  optimal mapping  that minimizes   asymptotic error rate   is not very relevant  for practical communication systems.  Moreover, finding the optimal mapping that minimizes error-floor is computationally expensive. Therefore, designing an efficient  MD mapping for BICM-ID that  offers   good BER in both  low and high SNR regions is very desirable. 
Motivated by the above discussion,  our objective  is to develop  an efficient MD mapping method  for BICM-ID to achieve   good BER performance   over AWGN and Rayleigh fading channels in  both low and high SNR regions. In what follows, we investigate     the  design criteria to achieve a good  mapping for AWGN and Rayleigh fading channels. It is important to  mention  that in order to investigate the design criteria of a suitable   mapping in low SNR region, we consider BICM-ID performance at the first iteration.  \color{black}

\subsection{AWGN Channel}

\subsubsection{Low SNR region}

Let ${\mb d} = \lbrace d_{1}, \cdots, d_{p}\rbrace$ be the set of all possible Euclidean distances between two signal points in $\mb{\chi}$ where $d_{i}<d_{j}$ if ${i}< {j}$ and $p$ depends on the constellation. For example, $p$ takes value  two and five, respectively,   for  2-{D} and 4-D QPSK. A larger  value of $d_{1}$ is desired to achieve a better BER performance of BICM over AWGN channel  \cite{BICM}.  Moreover, in order to achieve  a good asymptotic BER performance of  BICM  over AWGN channel, the value of $N_{min}$ should be as small as possible \cite{BICM} where  $N_{min}$ is defined as
\begin{equation}
N_{min} = \dfrac{1}{mN2^{mN}}\sum_{i=1}^{mN}\sum_{b=0}^{1} \sum_{{\mb x}\in \mb{\chi}_{b}^{i}}  N({\mb x},i)
\label{Nmin}
\end{equation}  where   $\mb{x} = (x_{1}, x_{2}, \cdots, x_{N})$ is a $2N$-D signal point, $\mb{\chi}_{b}^{i}$ is the subset of all $\mb{x} \in \mb{\chi}$ whose labels take value $b$ in $i$th bit position,  and \color{black} $N({\mb x},i)$ is the number of signal points at the Euclidian \color{black}  distance $d_{1}$ from ${\mb x}$ that are different from ${\mb x}$ in  $i$th bit position.

\subsubsection{High SNR region}
 $\hat{d}_{min}^{2}$, which is defined as the minimum squared Euclidean distance (MSED) between two symbol-vectors with Hamming distance one,  is the dominant factor for the asymptotic performance of BICM-ID over AWGN channels \cite{Chindapol_16QAM}. \color{black}

\subsection{  Block-Fading \color{black} Channel}

 The so called \emph{harmonic mean}  of the MSED \cite{BICM} of mappings is a well-known parameter that relates  to the BER  performance of  BICM-ID in Rayleigh fading channels  and is defined as \cite{BICM} \color{black}
\begin{equation}
\Phi_{br}(\mu, \mb{\chi}) = \left(\dfrac{1}{mN2^{mN}}\sum_{i=1}^{mN} \sum_{b=0}^{1} \sum_{{\mb x}\in \mb{\chi}_{b}^{i}} \dfrac{1}{\Vert {\mb x}- \hat{\mb x} \Vert^{2}}\right) ^{-1}.
\label{Harmonic_fading}
\end{equation} 

 For the performance at the first iteration, $\hat{\mb x}$ refers to the nearest neighbor of $\mb{x}$ in $\mb{\chi}_{\bar{b}}^{i}$ and (\ref{Harmonic_fading}) is referred to  as \emph{harmonic mean} of the MSED before feedback. For the asymptotic performance $\mb{\chi}_{\bar{b}}^{i}$ involves only one symbol-vector $\hat{\mb{x}}$ which is different from $\mb{x}$ in only $i$th bit position \cite{Chindapol_16QAM}. In this case, (\ref{Harmonic_fading}) is referred to as \emph{harmonic mean} of the MSED after feedback which is denoted by $\hat{\Phi}_{br}(\mu, \mb{\chi})$\color{black}.
 To achieve  a good performance at the first iteration, a larger value of $\Phi_{br}(\mu, \mb{\chi})$ \color{black} is desired. Let $n_{i}$ be defined as
\begin{equation}
n_{i} = \sum_{j=1}^{mN} \sum_{b=0}^{1} \sum_{{\mb x}\in \mb{\chi}_{b}^{j}} I_{i}({\mb x}, \hat{\mb x}); ~~~~~~i = 1, \cdots, p,
\label{hat_n_i}
\end{equation} where   $I_{i}({\mb x}, \hat{\mb x})$ is an indicator function that takes value one if the Euclidean distance between ${\mb x}$ and $\hat{\mb x}$ is equal to $d_{i}$, otherwise it is equal to zero. Then, substituting  (\ref{hat_n_i}) into (\ref{Harmonic_fading}), we  obtain
\begin{equation}
\Phi_{br}(\mu, \mb{\chi}) = \left( \dfrac{1}{mN2^{mN}}\sum_{i=1}^{p} \dfrac{n_{i}}{d_{i}^{2}}\right) ^{-1}.
\label{Rewritten_Harmonic_fading}
\end{equation}\color{black}

In what follows, we describe  the mapping design criteria  for   block-fading \color{black} channel in low and high SNR regions. 

\subsubsection{Low SNR region}  In  low SNR region,   $\hat{\mb{x}}$ in (\ref{Harmonic_fading}) is the nearest neighbour of $\mb{x}$ in $\mb{\chi}_{\bar b}^{i}$. Since each signal is different from its nearest neighbor at least in one bit position,  $\hat{\mb{x}}$ is at the Euclidean distance $d_{1}$ from $\mb{x}$ for some values of $i$. As a result, $n_{1}$ in (\ref{Rewritten_Harmonic_fading}) has a non-zero value. Moreover, summation of $n_{i}$ for all  $i$ is  constant,   i.e., 
\begin{equation}
\sum_{i = 1}^{p} n_{i}=mN2^{mN}.
\label{sigma_n_i}
\end{equation} 
Considering (\ref{Rewritten_Harmonic_fading}) and (\ref{sigma_n_i}), it is obvious that  any reduction in $n_{i}$ for a specific $i$ without increasing $n_{j}$, where $j<i$, yields a larger value of  $\Phi_{br}(\mu,\mb{\chi})$.
In particular, one can  increase  $\Phi_{br}(\mu,\mb{\chi})$ by  decreasing  $n_1$. 

\subsubsection{High SNR region}
For the asymptotic performance of BICM-ID, $\hat{\mb{x}}$ is considered to be  different from $\mb{x}$ only in $i$th bit position. In other words,   $\mb{\chi}_{\bar b}^{i}$ includes only one signal. 
\color{black} As a result, it is possible to design a mapping in which $\hat{d}_{min}>d_{i}$ for some small values of $i$. This gives  $n_{i}=0$ for some small values of $i$ and it  yields a greater value for  $\hat{\Phi}_{br}(\mu,\mb{\chi})$\color{black}. Numerical examples \color{black} show that a significant increase in $\hat{d}_{min}$ leads to a considerable enhancement in $\hat{\Phi}_{br}(\mu,\mb{\chi})$.  


\subsection{Fast Rayleigh Fading Channel}
The effect of mapping  on the BER performance of BICM-ID over the fast Rayleigh fading channel is characterized by $\Phi_{fr}(\mu, \mb{\chi})$ which is expressed as  \cite{hypercub_conf} \color{black}
\begin{equation} 
\Phi_{fr}(\mu, \mb{\chi}) = \left( \dfrac{1}{mN2^{mN}}\sum_{i=1}^{mN} \sum_{b=0}^{1} \sum_{{\mb x}\in \mb{\chi}_{b}^{i}} \prod_{j = 1}^{N} \left( 1 + \dfrac{\Vert {x_{j}}- \hat{x}_{j} \Vert^{2}}{4N_{0}}\right) ^{-1}\right) ^{-1}.
\end{equation} In particular, a   greater value of $\Phi_{fr}$ offers a better BER performance for BICM-ID. Similar to the   block-fading \color{black} case, it is easy to show that by decreasing $n_{1}$ and increasing $\hat{d}_{min}$ one can improve $\Phi_{fr}(\mu, \mb{\chi})$ at low and high SNR values, respectively.

\color{black}

In summary, it can be concluded that a mapping which offers a small value of $N_{min}$ while it gives a large value of $\hat{d}_{min}$, is suitable to achieve good error performance of BICM-ID  over both AWGN and Rayleigh fading channels in low and high  SNR regions. A small value of $N_{min}$ implies that the average Hamming distance between neighbouring  symbols, i.e., symbols with Euclidean distance $d_{1}$, is small and consequently  $n_{1}$ is small. This eventually increases $\Phi_{br}(\mu, \mb{\chi})$ and $\Phi_{fr}(\mu, \mb{\chi})$ at the first iteration of BICM-ID in Rayleigh fading channels. So, a small value   of $N_{min}$ can improve the mapping's performance at the first iteration, i.e., in low SNR region not only in AWGN channel but also in Rayleigh fading channels. 
On the other hand, a large value of $\hat{d}_{min}$ implies that $n_{i}$ is equal to zero for small values of $i$. This increases  $\hat{\Phi}_{br}(\mu, \mb{\chi})$ and $\Phi_{fr}(\mu, \mb{\chi})$ at high SNR values.  As a consequence, the asymptotic BER performance of BICM-ID improves over  AWGN and Rayleigh fading channels as the value of   $\hat{d}_{min}$ increases. 

\section{Proposed Mapping Method}
\label{proposed_map}
Based on the above discussion, we take a heuristic approach to obtain a mapping that offers good BER performance in low and high  SNR regions  over AWGN and Rayleigh fading channels. In particular,  we apply two key techniques  as follows. First, to generate a mapping with a large value of $\hat{d}_{min}$, we map binary labels with the  Hamming distance one to the symbol-vectors with a large Euclidean distance. This leads to  good error-floors in  Rayleigh fading and AWGN channels. Second,  most of the nearest neighbouring  symbol-vectors are mapped to  the binary labels that have  Hamming distance two. This results in a small value of $N_{min}$ and yields  good BER performance in   low SNR region over AWGN and Rayleigh fading channels. 

The proposed MD mapping using  $2^{m}$-QAM symbols 
is obtained progressively in $(m-1)$ steps. The mappings in steps $i~(1\leq i \leq (m-2))$ are   intermediate mappings whereas the mapping in step $i=(m-1)$ is the final  mapping.  In   $i$th step,  $2^{i+1}$ symbols from $2^{m}$-ary constellation are selected to be  used in the mapping process. 
The symbols selection process is descried as below. 
\subsection{Symbols Selection Process}
 We assume that position-indexes of symbols in a given QAM constellation,  increase by moving right or down.  For example, Fig. \ref{16QAM_steps}(a) shows   a $16$-QAM constellation   where $S_{j}$ represents the symbol with position-index $j$ and $j$ increases from left to right or  from top to bottom. 
  The \emph{general-principles} in choosing $2^{i+1}$ symbols from a $2^{m}$-QAM constellation in  $i$th step are as follows: (i) by moving the set  of  selected symbols one can  cover all symbols of the constellation, provided that each symbol is covered only one time.  In other words, the  square $M$-QAM constellations can be partitioned into a number of  subsets where the structures/shapes formed by these subsets are congruent with one another. Thus, by moving one of the subsets and superimposing it on the remaining subsets, one can cover all symbols in the constellation while each symbol is covered only once. \color{black} (ii) The MSED between the chosen symbols is as large as possible.  Without loss of   generality, let us use ${\chi}_{i}$ to  denote the set of  $2^{i+1}$ chosen symbols in step $i$ and ${\mb \alpha}_{i}=[\alpha^{(1)}_{i}, \alpha^{(2)}_{i}, \cdots, \alpha^{(2^{i+1})}_{i}]$  indicates the position-indexes of symbols in ${\chi}_{i}$.  The set of used symbols in step $(i+1)$ contains all the used symbols in step $i$, i.e., ${\chi}_{i} \subset {\chi}_{i+1}$ and ${\mb \alpha}_{i} \subset {\mb \alpha}_{i+1}$. 
  \begin{example}
In step $i = 1$,  four symbols are selected from a 16-QAM constellation to be used  in the mapping process. 
There are four distinct sets, which are made of four symbols,  as follows:    $\mathcal{S}_{1,1}=\lbrace S_{1}, S_{2},S_{3},S_{4} \rbrace$, $\mathcal{S}_{1,2}=\lbrace S_{1}, S_{2},S_{5},S_{6} \rbrace$, $\mathcal{S}_{1,3}=\lbrace S_{1}, S_{3},S_{5},S_{7} \rbrace$, and $\mathcal{S}_{1,4}=\lbrace S_{1}, S_{3}, S_{9}, S_{11} \rbrace$. It is obvious that by moving  any of these sets, one can cover all symbols of 16-QAM constellation, provided that each symbol is covered only once. 
Among these sets, $\mathcal{S}_{1,4}$ provides the maximum value of MSED between its symbols. As a result, for 16-QAM, we have ${\chi}_1=\mathcal{S}_{1,4}$ and $\mb{\alpha}_{1} = [1, 3, 9, 11]$. In Fig. \ref{16QAM_steps}(b) dark symbols indicate the symbols in $\mathcal{\chi}_{1}$. 

\begin{figure*}
\centering
\includegraphics[width= 0.8\columnwidth, viewport=30mm 210.94mm 152mm 254.00mm]{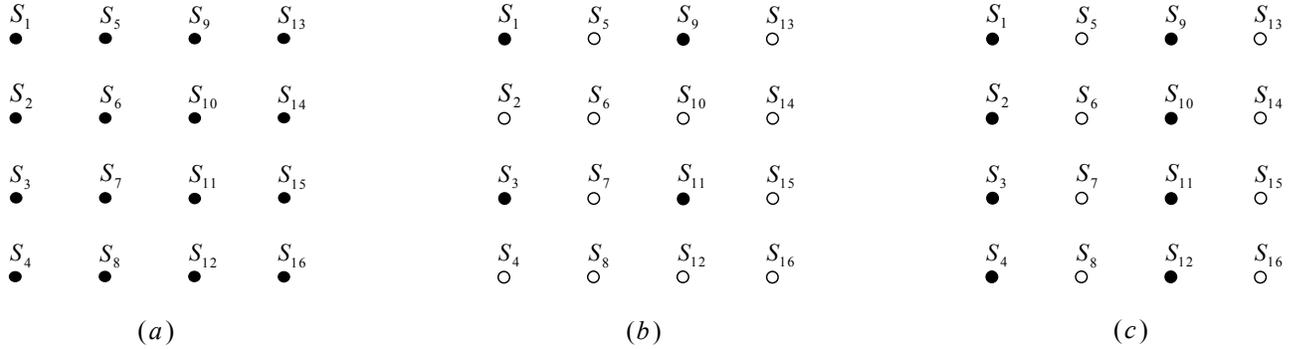} 
\caption{(a)  A 16-QAM constellation, (b) Four selected  16-QAM symbols (dark symbols) to be used in  our mapping in  step $i=1$, and  (c) Eight selected 16-QAM symbols  (dark symbols) to be used in our mapping  in step $i=2$.}
\label{16QAM_steps}
\end{figure*}
  
In step $i= 2$, eight   symbols are selected and  these eight symbols should   include all symbols in $\chi_{1}$. There are two sets of eight symbols as follows:  $\mathcal{S}_{2,1}=\lbrace S_{1}, S_{2},S_{3},S_{4},S_{9}, S_{10},S_{11},S_{12} \rbrace$ and $\mathcal{S}_{2,2}=\lbrace S_{1}, S_{3},S_{5},S_{7},S_{9}, S_{11},S_{13},S_{15} \rbrace$,  that include all symbols in $\chi_{1}$, and also can cover all the symbols in the constellation by moving right or down. Both $\mathcal{S}_{2,1}$ and $\mathcal{S}_{2,2}$ offer the same  MSED between the selected symbols. Therefore, either $\mathcal{S}_{2,1}$ or $\mathcal{S}_{2,2}$  can be selected in step 2.  We assume that the set of the selected 16-QAM symbols in the second step,  ${\chi}_{2}= \mathcal{S}_{2,1}$. The symbols in ${\chi}_2$  are shown by dark colour  in Fig. \ref{16QAM_steps}(c). 
The corresponding vector of position-indexes for $\chi_2$  is $\mb{\alpha}_{2} = [1, 2, 3, 4, 9, 10, 11, 12]$. In step $i=3$, all 16-QAM symbols are used  in our  mapping process. 
\end{example} \color{black}

\subsection{Mapping Process}

 Let ${\mb l}  = [l^{(1)}, l^{(2)}, \cdots, l^{(mN)}]$ be a $mN$-bit binary label and in step $i$, ${\mb a}_{i}=[ a^{(1)}_{i}, a^{(2)}_{i}, \cdots, a^{((i+1)N)}_{i}]$ denotes $(i+1)N$ least significant bits of ${\mb l}$ where $a^{(k)}_{i}$ is given by
\begin{equation}
a^{(k)}_{i} = l^{(mN-(i+1)N+k)},~~~~~~~~~~k=1,2,\cdots, (i+1)N.
\end{equation} 
%
Let ${\mb a}_{i}$ be mapped to symbol-vector ${\mb x}_{i}=[x^{(1)}_{i}, \cdots, x^{(N)}_{i}]$ where $x^{(k)}_{i}\in {\chi}_i$. We denote the corresponding position-index vector for ${\mb x}_{i}$ by ${\mb j}_{i}=[j^{(1)}_{i}, \cdots, j^{(N)}_{i}]$, where $j^{(k)}_{i}\in {\mb \alpha}_{i}$ refers to the position-index of symbol $x^{(k)}_{i}$.   
The steps of our  proposed mapping method are described in the following. 
\subsubsection{First step} 
In step $i=1$,  the selected symbol set $\chi_1$   is  equivalent to  QPSK symbols in terms of intersymbol  Euclidean distances. 
 Therefore, in order to achieve a good mapping, we apply the efficient MD QPSK mapping method introduced in \cite{MD-BQPSK-Simoen}. In particular,  t\color{black} he $2N$-bit label, $\mb{a}_{1}$, is mapped to $N$ consecutive QPSK symbols using the method proposed in \cite{MD-BQPSK-Simoen}. \color{black} Then, we use a  conversion vector, denoted by $\mb{\gamma} = [\gamma^{(1)}, \cdots, \gamma^{(4)}]$, to convert each symbol in the achieved MD QPSK mapping to one of symbols in $\chi{_1}$. Table \ref{alpha_gamma} provides the proposed conversion vectors. 
 
 \begin{table}[h!]
\caption{Conversion vector, $\mb{\gamma}$.}
\centering
\begin{tabular}{|c||c|} 
\hline
Basic Modulation & $\mb{\gamma}$ \\ \hline \hline
$16$-QAM &  [11, 3, 1,  9]  \\ \hline
$64$-QAM & [37, 5, 1, 33]\\ \hline
\end{tabular}
\label{alpha_gamma}
\end{table}
 
 The conversion process is described as follows. Without loss of generality, we assume the QPSK symbols are expressed  as 
\begin{equation}
P_{k}= e^{j\frac{\pi k}{2}}; ~~~ k = 1, \cdots, 4; ~~~ j^{2} = -1,
\end{equation} 
where, $k$ is the symbol position-index in QPSK constellation.   A particular QPSK symbol $P_{k}$ is converted to one of symbols in $\chi_1$ as follows 
\begin{equation}
\label{convert}
P_{k}\rightarrow S_{z}; ~~~ z = \gamma^{(k)},
\end{equation}
where $S_{z}$ is the symbol with position-index $z$ in $2^{m}$-QAM constellation. 
 It is important to note that   $\mb{\gamma}$, converts each QPSK symbol  to the corresponding symbol in the 4-ary constellation created using  the four  chosen $2^{m}$-QAM symbols. As a consequence, all properties  \color{black}of the  MD QPSK mapping  \cite{MD-BQPSK-Simoen} are conserved for our MD mapping using  four  selected $2^m$-QAM symbols.
\begin{example}
\label{first_step}
Let in our  proposed MD mapping method, $m = 4$ (16-QAM), $N = 2$, and $\mb{l} = [1,1,1,0,0,1,1,1]$.  For this example, $\mb{a}_{1}$   is made of four    least significant bits of ${\mb l}$, 
i.e., 
$\mb{a}_{1} = [0,1,1,1]$.
 In step $i=1$, $\mb{a}_{1}$ is mapped to a vector of two QPSK symbols following the proposed method in \cite{MD-BQPSK-Simoen} that results  QPSK symbol-vector  $\mb{P} = [P_{4},P_{2}]$. Then  $\mb{P}$ is converted to the 16-QAM symbol-vector $\mb{x}_{1}$  using  $\mb{\gamma}$.  Therefore, by applying (\ref{convert}) and $\mb{\gamma} = [11, 3, 1, 9]$ (c.f., Table \ref{alpha_gamma}   for 16-QAM),  $\mb{x}_{1}$  is obtained as
\begin{eqnarray}
& \mb{x}_{1} = [S_{{\gamma}^{(4)}}, S_{{\gamma}^{(2)}}] = [S_{9},S_{3}].
\end{eqnarray}  The vector of position-indexes corresponding to   the symbol-vector $\mb{x}_{1}$, $\mb{j}_{1}=[9,3]$. 
\end{example} 
 






\subsubsection{Subsequent steps} 
 In general, in step ~$i~ ( i=2,3\cdots,  m-1$),  label ${\mb a}_{i}$ is mapped to a vector of $N$ symbols   using the intermediate  mapping in  step $(i-1)$ and symbol set $\chi_i$. \color{black}     Let in step $i$, ${\mb b}_{i}=[b^{(1)}_{i}, b^{(2)}_{i}, \cdots, b^{(N)}_{i}]$ denote the $N$  most significant \color{black} bits of $\mb{a}_{i}$, i.e., $b_{i}^{(k)} = a_{i}^{(k)}$ for $k = 1, \cdots, N$. Each symbol in  $\mb{x}_{i-1}$ is transformed to obtain  symbol-vector in step $i$, $\mb{x}_{i}$.   The transformation rule is defined by   $\mb{\beta}_{i,k}$, i.e., 
 \begin{equation}
\label{convert2}
x_{i-1}^{(k)}\xrightarrow {\mb{\beta}_{i,k}}x_{i}^{(k)}, ~~~x_{i-1}^{(k)}\in \chi_{i-1},~~~x_{i}^{(k)} \in \chi_{i},
\end{equation}
 where  $\mb{\beta}_{i,k}$ 
 is determined based on  the Hamming weight of ${\mb b}_{i}$ and bit value of  $b_{i}^{(k)}$. 
 
 \emph{ Design consideration   of ${\mb \beta}_{i,k}$:}
 There are two key ideas in designing $\mb{\beta}_{i,k}$ ($i>1$) as follows. As discussed in Section \ref{Guide-Lines}, in order to achieve a good error performance at high SNRs, a larger value of $\hat{d}_{min}^{2}$ is desired for AWGN and Rayleigh fading channels. Let $\hat{d}_{min,i}^{2}$ be the MSED between two symbol-vectors with Hamming distance one in  $i$th step of our  proposed mapping process. As mentioned earlier that the  intermediate  MD mapping in  first step is  equivalent to the optimum MD QPSK mapping developed in \cite{MD-BQPSK-Simoen}. Therefore, it  yields the largest possible value of  $\hat{d}_{min,1}^{2}$ for the selected four symbols from $2^m$-QAM. In order to  achieve a large  value of  $\hat{d}_{min}^{2}$,  $\mb{\beta}_{i,k}$ should be designed such that  $\hat{d}_{min,i}^{2} \geq \hat{d}_{min,1}^{2}$ for $i=2,3,\cdots, m-1$. As it is mentioned earlier, it is desirable to achieve a  value of  the average Hamming distance between the nearest symbol-vectors, i.e., $N_{min}$, close to two. To meet this goal,  $\mb{\beta}_{i,k}$ is designed such that  the most of the symbol-vectors with the Euclidean distance $d_{min,i}$ are mapped by binary labels with Hamming distance two in each step;  where $d_{min,i}$ is the minimum Euclidean distance between the symbols in $\chi_{i}$\color{black}. 
Based on the above discussion, we design $\mb{\beta}_{i,k}$ as follows.

Let $\mb{a}_{i} = [\mb{b}_{i},\mb{a}_{i-1}]$ be a given label in step $i$ where $\mb{b}_{i}$ and $\mb{a}_{i-1}$ are two binary sequences of lengths  $N$ and $iN$ bits, respectively. Assume that $\hat{\mb{a}}_{i}$ is a binary sequence of $(i+1)N$ bits and is different from $\mb{a}_{i}$ only in $k$th bit position. Then, there are two possible cases for $\hat{\mb{a}}_{i}$ as follows
\begin{equation}
\label{a_hat}
\hat{\mb{a}}_{i}  =\left\{
\begin{array}{ll}
[\hat{\mb{b}}_{i},\mb{a}_{i-1}] & \mbox{if}~ k\leq N,\\      

[\mb{b}_{i},\hat{\mb{a}}_{i-1}] & \mbox{if}~ k>N,\\ 
\end{array}\right.
\end{equation} where $\hat{\mb{b}}_{i}$ and $\hat{\mb{a}}_{i-1}$ are  $N$ and $iN$ bit sequences that 
have Hamming distance one from  $\mb{b}_{i}$ and $\mb{a}_{i-1}$, respectively. Our first goal is to map  $\mb{a}_{i}$ and $\hat{\mb{a}}_{i}$ to the symbol-vectors $\mb{x}_{i}$ and $\hat{\mb{x}}_{i}$, respectively such that 
\begin{equation}
\label{first_goals}
\Vert \mb{x}_{i} - \hat{\mb{x}}_{i}\Vert ^{2} \geqslant \hat{d}_{min,1}^{2}.
\end{equation} 
 Let $\tilde{\mb{a}}_{i}$ be a binary sequence of length $(i+1)N$ and different from $\mb{a}_{i}$ in $j$th and $k$th bit positions where $j < k \leqslant (i+1)N$. Then, $\tilde{\mb{a}}_{i}$ can be defined as one of the three following possible cases
\begin{equation}
\label{a_tilde}
\tilde{\mb{a}}_{i}  =\left\{
\begin{array}{ll}
[\tilde{\mb{b}}_{i},\mb{a}_{i-1}] & \mbox{if}~ j < N,k \leqslant N,\\ 

[\hat{\mb{b}}_{i},\hat{\mb{a}}_{i-1}] & \mbox{if}~ j\leqslant N,k>N,\\ 

[\mb{b}_{i},\tilde{\mb{a}}_{i-1}] & \mbox{if}~ j> N,k>N, \\ 
\end{array}\right.
\end{equation}
where $\tilde{\mb{b}}_{i}$ is a $N$-bit sequence with Hamming distance two from  $\mb{b}_{i}$ and  $\tilde{\mb{a}}_{i-1}$ is a $iN$-bit sequence with Hamming distance two from $\mb{a}_{i-1}$. Suppose that in step $i$, $\tilde{\mb{x}}_{i} \in \mb{\psi}_{i}$, where $\mb{\psi}_{i}$ denotes the set of the nearest symbol-vectors to $\mb{x}_{i}$ in $\mb{\chi}_i=\chi_i^N$.
Our second goal is to map the most of the symbol-vectors in $\mb{\psi}_{i}$ by one of the possible cases of $\tilde{\mb{a}}_{i}$ in (\ref{a_tilde}). As such  we can have 
\begin{equation}
\label{second-goals}
d_{H}(\mb{x}_{i},\tilde{\mb{x}}_{i}) = 2,
\end{equation} for the most cases of $\tilde{\mb{x}}_{i}$, where $d_{H}(a,b)$ denotes the 
Hamming distance between $a$ and $b$. \color{black}

Above mentioned two  goals are achieved via a systematic    symbol transformation    from step $(i-1)$ to step $i$ using $\mb{\beta}_{i,k}$. 
 Specifically, $\mb{\beta}_{i,k}$
 depends on the Hamming weight  of $\mb{b}_{i}$  as well as the bit value ${b}_{i}^{(k)}$.  So, there can be  four possible cases  which are  represented by  $\mb{\beta}_{E0}$, $\mb{\beta}_{E1}$, $\mb{\beta}_{O0}$, and $\mb{\beta}_{O1}$ as follows  
 
\begin{equation}
\label{beta}
\mb{\beta}_{i,k}  =\left\{
\begin{array}{ll}
\mb{\beta}_{E0} & \mbox{if}~ w_{H}(\mb{b}_{i})\in \mathbb{E},~b^{(k)}_{i}=0, \\
\mb{\beta}_{E1} & \mbox{if}~ w_{H}(\mb{b}_{i})\in \mathbb{E},~b^{(k)}_{i}=1, \\
\mb{\beta}_{O0} & \mbox{if}~ w_{H}(\mb{b}_{i})\in \mathbb{O},~b^{(k)}_{i}=0, \\
\mb{\beta}_{O1} & \mbox{if}~ w_{H}(\mb{b}_{i})\in \mathbb{O},~b^{(k)}_{i}=1, \\
\end{array}\right.
\end{equation}
where 
$\mb{\beta}_{E0}$, $\mb{\beta}_{E1}$, $\mb{\beta}_{O0}$, and $\mb{\beta}_{O1}$ are vectors of the position-indexes of  symbols in $\chi_i$.  
Tables \ref{Indx_vct_16QAM} and \ref{Indx_vct_64QAM} provide our proposed vectors    $\mb{\beta}_{E0}$, $\mb{\beta}_{E1}$, $\mb{\beta}_{O0}$, and $\mb{\beta}_{O1}$ for different steps in the MD mapping using  16-QAM and 64-QAM, respectively.

\begin{table}[h!]
\caption{$\mb{\alpha}_{i-1}$ and different $\mb{\beta_{i,k}}$ for 16-QAM.}
\centering
\resizebox{0.4\columnwidth}{!}{%
\begin{tabular}{|c||c|c|}
\hline
Index-vector & $i = 2$ & $i = 3$ \\ \hline \hline
$\mb{\alpha}_{i-1}$ & [1, 3, 9, 11] & [1, 2, 3, 4, 9, 10, 11, 12] \\ \hline
$\mb{\beta}_{E0}$ & [1, 3, 9, 11] & [1, 2, 3, 4, 9, 10, 11, 12] \\ \hline
$\mb{\beta}_{E1}$ & [2, 4, 10, 12] & [5, 6, 7, 8, 13, 14, 15, 16] \\ \hline
$\mb{\beta}_{O0}$ & [11, 9, 3, 1] & [11, 12, 9, 10, 3, 4, 1, 2]\\ \hline
$\mb{\beta}_{O1}$ & [12, 10, 4, 2] & [15, 16, 13, 14, 7, 8, 5, 6] \\ \hline
\end{tabular}
\label{Indx_vct_16QAM}
}
\end{table}

\begin{table*}[h!]
\caption{$\mb{\alpha}_{i-1}$ and different $\mb{\beta_{i,k}}$ for 64-QAM.}
\centering
\resizebox{1\columnwidth}{!}{%
\begin{tabular}{|c||c|c|c|c|}
\hline
Index-vector & $i = 2$ & $i = 3$ & $i = 4$ & $i = 5$\\ \hline \hline
\multirow{2}{*}{$\mb{\alpha}_{i-1}$} & \multirow{2}{*}{[1, 5, 33, 37]} & \multirow{2}{*}{[1, 3, 5, 7, 33, 35, 37, 39]} & \multirow{2}{*}{[1, 3, 5, 7, 17, 19, 21, 23, 33, 35, 37, 39, 49, 51, 53, 55]} & [1, 3, 5, 7, 9, 11, 13, 15, 17, 19, 21, 23, 25, 27, 29, 31, 33,  \\ &&&& 35, 37, 39, 41, 43, 45, 47, 49, 51, 53, 55, 57, 59, 61, 63] \\ \hline

\multirow{2}{*}{$\mb{\beta}_{E0}$} & \multirow{2}{*}{[1, 5, 33, 37]} & \multirow{2}{*}{[1, 3, 5, 7, 33, 35, 37, 39]} & \multirow{2}{*}{[1, 3, 5, 7, 17, 19, 21, 23, 33, 35, 37, 39, 49, 51, 53, 55]} &   [1, 3, 5, 7, 9, 11, 13, 15, 17, 19, 21, 23, 25, 27, 29, 31, 33, \\ &&&& 35, 37, 39, 41, 43, 45, 47, 49, 51, 53, 55, 57, 59, 61, 63] \\ \hline

\multirow{2}{*}{$\mb{\beta}_{E1}$ } & \multirow{2}{*}{[3, 7, 35, 39]} & \multirow{2}{*}{[17, 19, 21, 23, 49, 51, 53, 55]} & \multirow{2}{*}{[9, 11, 13, 15, 25, 27, 29, 31, 41, 43, 45, 47, 57, 59, 61, 63]} &  [2, 4, 6, 8, 10, 12, 14, 16, 18, 20, 22, 24, 26, 28, 30, 32, 34,\\ &&&&   36, 38, 40, 42, 44, 46, 48, 50, 52, 54, 56, 58, 60, 62, 64] \\ \hline

\multirow{2}{*}{$\mb{\beta}_{O0}$} & \multirow{2}{*}{[37, 33, 5, 1]} & \multirow{2}{*}{[37, 39, 33, 35, 5, 7, 1, 3]} & \multirow{2}{*}{[37, 39, 33, 35, 53, 55, 49, 51, 5, 7, 1, 3, 21, 23, 17, 19]} &  [37, 39, 33, 35, 45, 47, 41, 43, 53, 55, 49, 51, 61, 63, 57,  \\ &&&& 59, 5, 7, 1, 3, 13, 15, 9, 11, 21, 23, 17, 19, 29, 31, 25, 27] \\ \hline

\multirow{2}{*}{$\mb{\beta}_{O1}$} & \multirow{2}{*}{[39, 35, 7, 3]} & \multirow{2}{*}{[53, 55, 49, 51, 21, 23, 17, 19]} & \multirow{2}{*}{[45, 47, 41, 43, 61, 63, 57, 59, 13, 15, 9, 11, 29, 31, 25, 27]} &  [38, 40, 34, 36, 46, 48, 42, 44, 54, 56, 50, 52, 62, 64, 58,\\ &&&&  60, 6, 8, 2, 4, 14, 16, 10, 12, 22, 24, 18,
20, 30, 32, 26, 28] \\ \hline

\end{tabular}
\label{Indx_vct_64QAM}
}
\end{table*}

\emph{Symbol transformation using $\mb{\beta}_{i,k}$:} For  given  $\mb{j}_{i-1}$ and $\mb{\alpha}_{i-1}$ and for a particular value of $k$ ($k = 1, \cdots, N$), there exists a $q \in \lbrace 1, \cdots, 2^{i}\rbrace$  such that 
$\label{beta_to_sym}
j^{(k)}_{i-1} = \alpha^{(q)}_{i-1}.$
Then the position-index of $k$th  symbol in  $\mb{x}_{i}$, i.e.,   $j^{(k)}_{i}$ is given by 
\begin{equation}
\label{beta_to_symbol}
j^{(k)}_{i} = \beta^{(q)}_{i,k},
\end{equation}
where $\beta^{(q)}_{i,k}$ is the $q$th element of the corresponding vector $\mb{\beta}_{i,k}$.   The values of $j^{(k)}_{i}$ determine the symbols in $\mb{x}_{i}$.   As an example,  Table  \ref{Transform_stp1_to_stp2} shows 
the transformation from $x_{1}^{(k)}$ to $x_{2}^{(k)}$ (using  $\mb{\beta}_{2,k}$  in     Table \ref{Indx_vct_16QAM}) and  Table \ref{Transform_stp2_to_stp3} illustrates the  transformation from $x_{2}^{(k)}$ to $x_{3}^{(k)}$ in   our  proposed MD mapping method using 16-QAM (using  $\mb{\beta}_{3,k}$  in     Table \ref{Indx_vct_16QAM}).
 


\begin{table}[h!]
\caption{Transformation of symbols from step $1$ to step $2$ using $16$-QAM.}
\centering
\resizebox{0.3\columnwidth}{!}{%
\begin{tabular}{|c||c|c|c|c|}
\hline
\multirow{2}{*}{$x_{1}^{(k)}$}&\multicolumn{4}{c|}{$x_{2}^{(k)}$}\\
\cline{2-5}
  & $\mb{\beta}_{E0}$ & $\mb{\beta}_{E1}$ & $\mb{\beta}_{O0}$ & $\mb{\beta}_{O1}$ \\
\hline\hline
$S_{1}$ &  $S_{1}$ & $S_{2}$ & $S_{11}$ & $S_{12}$  \\ \hline
$S_{3}$ &  $S_{3}$ & $S_{4}$ & $S_{9}$ & $S_{10}$  \\ \hline
$S_{9}$ &  $S_{9}$ & $S_{10}$ & $S_{3}$ & $S_{4}$  \\ \hline
$S_{11}$ &  $S_{11}$ & $S_{12}$ & $S_{1}$ & $S_{2}$  \\ \hline
\end{tabular}
\label{Transform_stp1_to_stp2}
}
\end{table}

%
%
%
%
%
%
%

\begin{table}[h!]
\caption{Transformation of symbols from step $2$ to step $3$ using $16$-QAM.}
\centering
\resizebox{0.3\columnwidth}{!}{%
\begin{tabular}{|c||c|c|c|c|}
\hline
\multirow{2}{*}{$x_{2}^{(k)}$}&\multicolumn{4}{c|}{$x_{3}^{(k)}$}\\
\cline{2-5}
  & $\mb{\beta}_{E0}$ & $\mb{\beta}_{E1}$ & $\mb{\beta}_{O0}$ & $\mb{\beta}_{O1}$ \\
\hline\hline
$S_{1}$ & $S_{1}$  & $S_{5}$  & $S_{11}$ & $S_{15}$ \\ \hline

$S_{2}$ & $S_{2}$  & $S_{6}$  & $S_{12}$  & $S_{16}$ \\ \hline

$S_{3}$ & $S_{3}$ & $S_{7}$ & $S_{9}$ &  $S_{13}$ \\ \hline

$S_{4}$ & $S_{4}$  & $S_{8}$ & $S_{10}$ &  $S_{14}$ \\ \hline

$S_{9}$ & $S_{9}$  & $S_{13}$  & $S_{3}$ & $S_{7}$ \\ \hline

$S_{10}$ & $S_{10}$  & $S_{14}$ &  $S_{4}$ & $S_{8}$ \\ \hline

$S_{11}$ & $S_{11}$ & $S_{15}$ &   $S_{1}$ & $S_{5}$ \\ \hline

$S_{12}$ & $S_{12}$  & $S_{16}$ &   $S_{2}$  & $S_{6}$ \\ \hline
\end{tabular}
\label{Transform_stp2_to_stp3}
}
\end{table}


\begin{example}
Using  example \ref{first_step} 
and for step $i=2$,  we have $\mb{a}_{2} = [1,0,0,1,1,1]$ and corresponding  $\mb{b}_{2} = [1,0]$.  Let us consider that $\mb{a}_{2}$ is mapped to  symbol-vector  $\mb{x}_{2} = [x_{2}^{(1)}, x_{2}^{(2)}]$  in this step.  
  Since $\mb{b}_{2}$ has an odd Hamming weight and $b_{2}^{(1)} = 1$, according to (\ref{beta}) $\mb{\beta}_{2,1} = \mb{\beta}_{O1}$.
From Table \ref{Indx_vct_16QAM}, we have  $\mb{\alpha}_{1} = [1, 3, 9, 11]$ and $\mb{\beta}_{O1} = [12, 10, 4, 2]$. Since    $\mb{j}_{1}=[9,3]$ (see Example \ref{first_step}),  we have $j_{1}^{(1)} = \alpha_{1}^{(q)}$ when  $q = 3$.  Using $q=3$ and   (\ref{beta_to_symbol}),  we obtain
$j_{2}^{(1)} = \mb{\beta}_{2,1}^{(3)} = 4$.
 Since  $b_{2}^{2}=0$, we have $\mb{\beta}_{2,2} = \mb{\beta}_{O0}$ where  $\mb{\beta}_{O0} = [11, 9, 3, 1]$ (c.f.,  Table III for $i = 2$). Moreover,  we have $j_{1}^{(2)} = \alpha_{1}^{(q)}$ when $q=2$.  Using $q=2$ and   (\ref{beta_to_symbol}), we obtain   
$j_{2}^{(2)}
= \mb{\beta}_{2,2}^{(2)} = 9$.
 As a result, $\mb{j}_{2} = [4,9]$ which means that  $\mb{x}_{1}$  will  be transformed to $\mb{x}_{2} = [S_{4}, S_{9}]$. In other words, $\mb{a}_{2}$ is mapped to $\mb{x}_{2} = [S_{4}, S_{9}]$ in step $i=2$.
 










In  step $i = 3$, $\mb{a}_{3}=[1,1,1,0,0,1,1,1]$  and  $\mb{b}_{3} = [1,1]$. The Hamming weight of $\mb{b}_{3}$ is even and both elements of $\mb{b}_{3}$ are equal to one. As a result, in order to determine the elements of $\mb{j}_{3} = [j_{3}^{(1)}, j_{3}^{(2)}]$ we set    $\mb{\beta}_{3,1}= \mb{\beta}_{E1}$ and $\mb{\beta}_{3,2}= \mb{\beta}_{E1}$.  From Table \ref{Indx_vct_16QAM} for $i = 3$,  we have $\mb{\beta}_{E1} = [5, 6, 7, 8, 13, 14, 15, 16]$ and $\mb{\alpha}_{2} = [1, 2, 3, 4, 9, 10, 11, 12]$. In addition, in step $i=2$ we have $\mb{j}_{2} = [4,9]$. It is obvious that  $j_{2}^{(1)} = \alpha_{2}^{(q)}$  when $q = 4$   and $j_{2}^{(2)} = \alpha_{2}^{(q)}$ when $q = 5$. By applying (\ref{beta_to_symbol}) we have 
$j_{3}^{(1)} = \beta_{E1}^{(4)} = 8$ and $j_{3}^{(2)}=\beta_{E1}^{(5)} = 13$. Consequently, $\mb{j}_{3} = [8,13]$ which means that $\mb{x}_{2}$  will  be transformed to $\mb{x}_{3} = [S_{8}, S_{13}]$. In other words, $\mb{l}$ is finally  mapped to symbol-vector $\mb{x}_{3}=[S_{8},S_{13}]$.
\end{example}

\begin{example}
Table \ref{16-QAM MD} illustrates the proposed 4-D mapping using 16-QAM symbols.
In this table, the decimal label in $(j+1,k+1)$th entry is mapped to  symbol-vector $\mb{x}=[S_{j},S_{k}]$.
For example, decimal label 231 corresponding to binary label  $\mb{l} = [1,1,1,0,1,1,1]$, is $(9,14)$th entry of Table IV which is  mapped to symbol-vector $\mb{x} = [S_{8},S_{13}]$ according to our proposed 4-D mapping using 16-QAM symbols. 

\begin{table*}[h!]
\caption{The resulted  $4$-D 16-QAM mapping.}
\centering
\resizebox{0.8\columnwidth}{!}{%
\begin{tabular}{|c||c|c|c|c|c|c|c|c|c|c|c|c|c|c|c|c|}
\hline
  & $S_{1}$  & $S_{2}$ & $S_{3}$  & $S_{4}$ & $S_{5}$  & $S_{6}$ & $S_{7}$  & $S_{8}$ & $S_{9}$  & $S_{10}$ & $S_{11}$  & $S_{12}$ & $S_{13}$  & $S_{14}$ & $S_{15}$  & $S_{16}$  \\ \hline \hline

$S_{1}$ & 0 & 17 & 14 & 31 & 65 & 80 & 79 & 94 & 3 & 18 & 13 & 28 & 66 & 83 & 76 & 93 \\ \hline

$S_{2}$ & 33 & 48 & 47 & 62 & 96 & 113 & 110 & 127 & 34 & 51 & 44 & 61 & 99 & 114 & 109 & 124 \\ \hline

$S_{3}$ & 5 & 20 & 11 & 26 & 68 & 85 & 74 & 91 & 6 & 23 & 8 & 25 & 71 & 86 & 73 & 88 \\ \hline

$S_{4}$ & 36 & 53 & 42 & 59 & 101 & 116 & 107 & 122 & 39 & 54 & 41 & 56 & 102 & 119 & 104 & 121 \\ \hline

$S_{5}$ & 129 & 144 & 143 & 158 & 192 & 209 & 206 & 223 & 130 & 147 & 140 & 157 & 195 & 210 & 205 & 220 \\ \hline

$S_{6}$ & 160 & 177 & 174 & 191 & 225 & 240 & 239 & 254 & 163 & 178 & 173 & 188 & 226 & 243 & 236 & 253 \\ \hline

$S_{7}$ & 132 & 149 & 138 & 155 & 197 & 212 & 203 & 218 & 135 & 150 & 137 & 152 & 198 & 215 & 200 & 217 \\ \hline

$S_{8}$ & 165 & 180 & 171 & 186 & 228 & 245 & 234 & 251 & 166 & 183 & 168 & 185 & 231 & 246 & 233 & 248 \\ \hline

$S_{9}$ & 9 & 24 & 7 & 22 & 72 & 89 & 70 & 87 & 10 & 27 & 4 & 21 & 75 & 90 & 69 & 84 \\ \hline

$S_{10}$ & 40 & 57 & 38 & 55 & 105 & 120 & 103 & 118 & 43 & 58 & 37 & 52 & 106 & 123 & 100 & 117 \\ \hline

$S_{11}$ & 12 & 29 & 2 & 19 & 77 & 92 & 67 & 82 & 15 & 30 & 1 & 16 & 78 & 95 & 64 & 81 \\ \hline

$S_{12}$ & 45 & 60 & 35 & 50 & 108 & 125 & 98 & 115 & 46 & 63 & 32 & 49 & 111 & 126 & 97 & 112 \\ \hline

$S_{13}$ & 136 & 153 & 134 & 151 & 201 & 216 & 199 & 214 & 139 & 154 & 133 & 148 & 202 & 219 & 196 & 213 \\ \hline

$S_{14}$ & 169 & 184 & 167 & 182 & 232 & 249 & 230 & 247 & 170 & 187 & 164 & 181 & 235 & 250 & 229 & 244 \\ \hline

$S_{15}$ & 141 & 156 & 131 & 146 & 204 & 221 & 194 & 211 & 142 & 159 & 128 & 145 & 207 & 222 & 193 & 208 \\ \hline

$S_{16}$ & 172 & 189 & 162 & 179 & 237 & 252 & 227 & 242 & 175 & 190 & 161 & 176 & 238 & 255 & 224 & 241 \\ \hline

\end{tabular}
\label{16-QAM MD}
}
\end{table*}   
\end{example}
\color{black}

\section{Numerical Results and Discussion}
\label{num_result}
In this section we provide some numerical examples   to demonstrate the performance and advantage of our proposed MD mapping for  BICM-ID systems. We compare our resulted  mappings with   random mappings and also with \color{black} the mappings that are optimized for the considered channel models  using the well-known BSA. According to the   BSA, a cost function is calculated
for each symbol in the constellation. The BSA starts with an initial mapping and  then finds  the
 symbol with the highest cost in the constellation and switches the label of this symbol with the label of another symbol.  As such   the total cost is reduced as much as possible \cite{BSA}. The BSA is  the best known computer search technique to find suitable
mappings for BICM-ID. However, it becomes intractable to obtain good mappings for MD modulations  with higher alphabet size e.g.,  $6$-D $64$-QAM due to computational time complexity.    The considered random mappings for each channel type,  in this section, are obtained by selecting the best mappings from  a large number of randomly generated mappings. \color{black} 
We consider a rate-$1/2$ convolutional code with the generator polynomial of $(13, 15)_{8}$.  An interleaver of length about $10000$ bits is used. 
  All BER curves are presented with seven iterations and \color{black}all  gains reported  in this section are  measured at BER  of $10^{-6}$. 

\subsection{Performance in AWGN Channel}
\begin{figure}[H]
\centering
\includegraphics[width= 0.6\columnwidth,viewport= 20mm 0mm 320mm 225mm,clip]{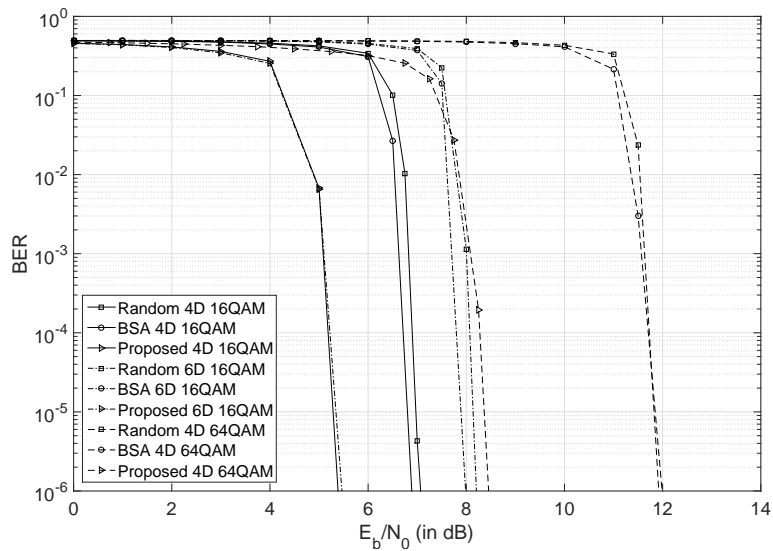}
\caption{BER performance in  AWGN channel.}
\label{BER_AWGN}
\end{figure}

\begin{figure}[H]
\centering
\includegraphics[width= 0.6\columnwidth,viewport= 20mm 0mm 320mm 225mm,clip]{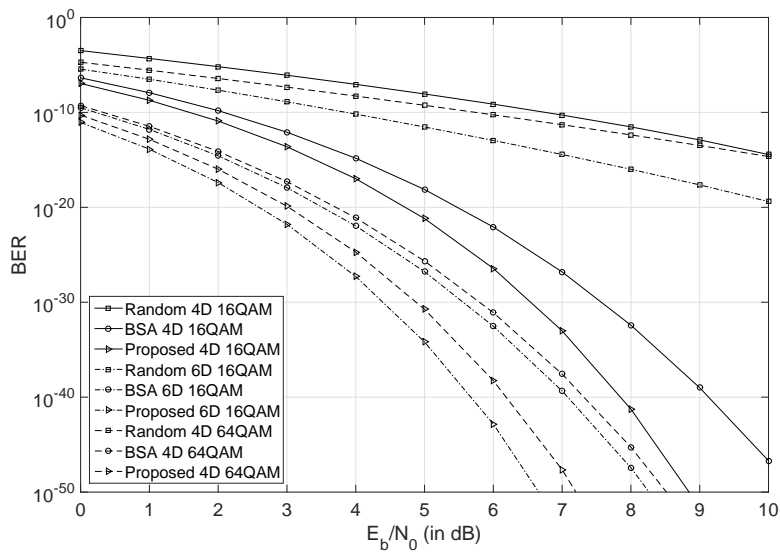}
\caption{Error-floor bounds in   AWGN channel.}
\label{EF_AWGN}
\end{figure}

As we mentioned earlier for AWGN channel, there are two important parameters for  a mapping,   i.e., $N_{min}$ and  $\hat{d}_{min}^{2}$ that are relevant  to  BER performance of BICM-ID systems.  In Table \ref{Parameters_AWGN}, we compare the values of $N_{min}$ and  $\hat{d}_{min}^{2}$ of our  MD mappings using $16$-QAM and $64$-QAM  with those of   well-known  BSA mappings   \color{black}  that are optimized for AWGN channel and  random mappings.    In this table,  BSA MD $64$-QAM   mapping  for higher dimension e.g., $N=3$ is not reported as it is not  obtained due to the computational complexity.   Table \ref{Parameters_AWGN} clearly shows that our mappings offer smaller values of $N_{min}$ compared to    their counterparts\color{black}.   So, our MD mappings  will improve the  BER performance of BICM-ID at low SNR  values over AWGN channel. This is confirmed in  Fig. \ref{BER_AWGN} which plots the BER of   various mappings \color{black} over  AWGN channel. As it can be observed from this figure, the proposed mappings outperform the BSA mappings by $1.5$ dB, $2.5$ dB and $3.5$ dB for $4$-D $16$-QAM, $6$-D $16$-QAM, and $4$-D $64$-QAM, respectively in low SNR region.   The improvement  with our proposed  mappings over  random mappings is even better.  \color{black} From Table \ref{Parameters_AWGN} it is also obvious that our mappings have  larger values of $\hat{d}_{min}^{2}$ compared to the BSA   and random mappings. Thus, the proposed \color{black} mappings   result  in  improved BER performance in high SNR region over AWGN channel.  This can be observed from the plotted   error-floor bounds    in Fig. \ref{EF_AWGN}.

\begin{table}[h!]
\caption{Comparison of $N_{min}$ and $\hat{d}_{min}^{2}$.}
\centering
\resizebox{0.7\columnwidth}{!}{%
\begin{tabular}{|c||c|c||c|c|}
\hline
\multirow{2}{*}{Mapping}&\multicolumn{2}{c||}{$N = 2$}&\multicolumn{2}{c|}{$N = 3$}\\
\cline{2-5}
  & $N_{min}$ & $\hat{d}_{min}^{2}$ & $N_{min}$ & $\hat{d}_{min}^{2}$ \\
\hline\hline

Random MD $16$-QAM & 4.04 & 0.2 &  6.01 & 0.1333  \\
\hline

BSA MD $16$-QAM &  3.7305 & 1.2 & 5.8254 & 1.3333  \\ \hline

Proposed MD $16$-QAM & 2.25 & 2.4 &  2.2778 & 2.6667  \\
 \thickhline
 
 Random MD $64$-QAM & 6.01 & 0.0476 &  8.9979 & 0.0317  \\
\hline

BSA MD $64$-QAM &  5.8240 & 1.1905 & -& - \\ \hline

Proposed MD $64$-QAM & 2.3214 & 2.2857 & 2.3571 & 2.5397  \\ \hline
\end{tabular}
\label{Parameters_AWGN}
}
\end{table}

\subsection{Performance in   Block-Fading \color{black}  Channel}
\begin{figure}[H]
\centering
\includegraphics[width= 0.6\columnwidth,viewport= 20mm 0mm 320mm 225mm,clip]{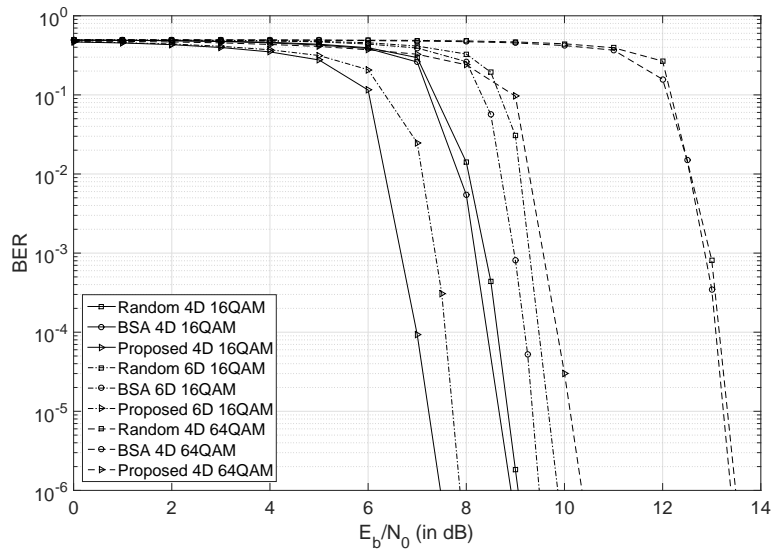}
\caption{BER performance in block-fading channel.}
\label{BER_Quasi-Rayleigh}
\end{figure}

\begin{figure}[H]
\centering
\includegraphics[width= 0.6\columnwidth,viewport= 20mm 0mm 320mm 225mm,clip]{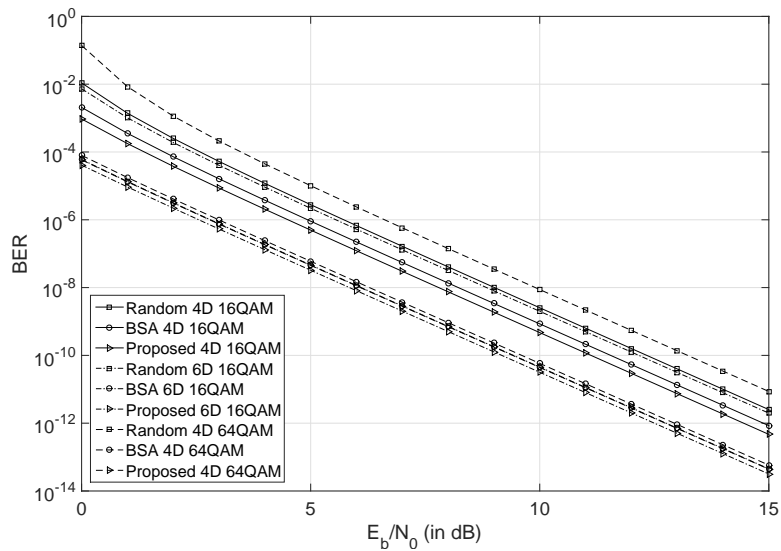}
\caption{Error-floor bounds in  block-fading  channel.}
\label{EF_Quasi_Rayleigh}
\end{figure}

For   block-fading \color{black} channel, $\Phi_{br}(\mu,\mb{\chi})$ and $\hat{\Phi}_{br}(\mu,\mb{\chi})$ are two important mapping parameters to compare    BER performance of BICM-ID systems. We compare the values of $\Phi_{br}(\mu,\mb{\chi})$ and $\hat{\Phi}_{br}(\mu,\mb{\chi})$ for various mappings in Table \ref{Parameters_QuasiFading}.  From  this table it is obvious that   our  mappings offer larger values of $\Phi_{br}(\mu,\mb{\chi})$ in comparison with the BSA   and random \color{black} mappings. This results in   better BER performance in low SNR region with our mappings. The  BER plots in
Fig. \ref{BER_Quasi-Rayleigh} show that  the proposed mappings offer a gain of  $1.5$ dB, $1.6$ dB and $3$ dB for $4$-D $16$-QAM, $6$-D $16$-QAM, and $4$-D $64$-QAM, respectively, compared to the BSA mappings that are optimized for   block-fading \color{black} channel.   The  performance gain with respect to  the random mappings is larger.  \color{black} From the  listed values of   $\hat{\Phi}_{br}(\mu,\mb{\chi})$  in Table \ref{Parameters_QuasiFading}, we  observe that  our  proposed mappings increase the values  of $\hat{\Phi}_{br}(\mu,\mb{\chi})$. Therefore,  our mappings offer  improved  error-floor  bounds as illustrated in Fig. \ref{EF_Quasi_Rayleigh}. 

\begin{table}[h!]
\caption{Comparison of $\Phi_{br}(\mu,\mb{\chi})$ and $\hat{\Phi}_{br}(\mu,\mb{\chi})$.}
\centering
\resizebox{0.7\columnwidth}{!}{%
\begin{tabular}{|c||c|c||c|c|}
\hline
\multirow{2}{*}{Mapping}&\multicolumn{2}{c||}{$N = 2$}&\multicolumn{2}{c|}{$N = 3$}\\
\cline{2-5}
  & $\Phi_{br}(\mu,\mb{\chi})$ & $\hat{\Phi}_{br}(\mu,\mb{\chi})$ & $\Phi_{br}(\mu,\mb{\chi})$ & $\hat{\Phi}_{br}(\mu,\mb{\chi})$ \\
\hline\hline

 Random MD $16$-QAM  &  0.2012 &   1.4350 & 0.1335 & 1.4934 \\ \hline
 
 BSA MD $16$-QAM  &  0.2026 &   2.5814 & 0.1342 & 2.8047 \\ \hline

Proposed MD 16-QAM & 0.2151 & 2.8491 & 0.1446 & 2.9741 \\ \thickhline

Random MD $64$-QAM  &  0.0478 & 1.1688 & 0.0318 & 1.4370 \\ \hline

BSA MD $64$-QAM  &  0.0481 & 2.6899 & - & - \\ \hline

Proposed MD $64$-QAM & 0.0579 & 2.8166 & 0.0392 & 2.9040  \\ \hline
\end{tabular}
\label{Parameters_QuasiFading}
}
\end{table}

\subsection{\color{black} Performance in Fast  Rayleigh Fading  Channel}
\begin{figure}[H]
\centering
\includegraphics[width= 0.6\columnwidth,viewport= 20mm 0mm 320mm 225mm,clip]{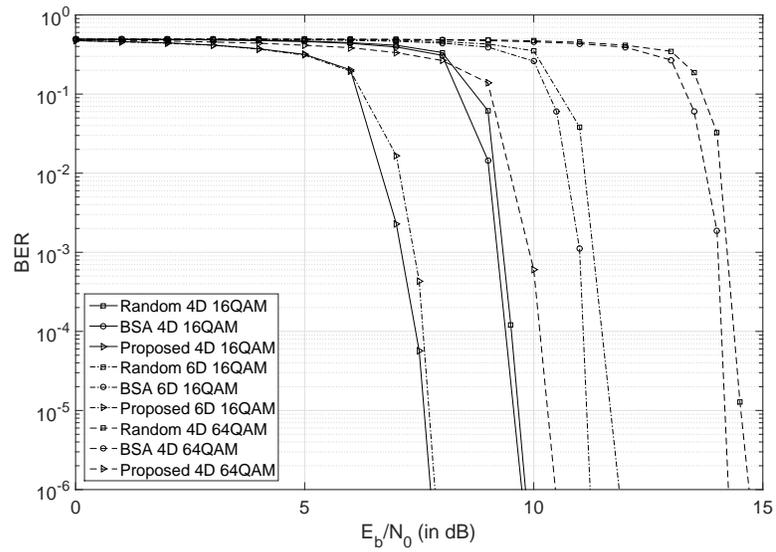}
\caption{BER performance in  fast Rayleigh fading channel.}
\label{BER_Fast_Rayleigh}
\end{figure}

\color{black} As mentioned earlier, $\Phi_{fr}(\mu,\mb{\chi})$ predicts  BER performance  of mappings  for  BICM-ID in  fast Rayleigh fading channel. Table \ref{Parameters_FastFading} compares the value of $\Phi_{fr}(\mu,\mb{\chi})$ for various mappings at different SNR values.  In order to calculate the values of $\Phi_{fr}(\mu,\mb{\chi})$ at low and high SNR values  we assume SNR$=2$ dB and SNR$=10$ dB, respectively. It is clear from  Table \ref{Parameters_FastFading} that our proposed mappings offer larger values of $\Phi_{fr}(\mu,\mb{\chi})$ at both low and high SNR values.  As a result, it is expected that our proposed mappings improve the BER in comparison with   their  counterparts. \color{black}  This is confirmed by the BER plots   in Fig. \ref{BER_Fast_Rayleigh}.  This figure shows that  our mappings outperform the BSA mappings by $2$ dB, $3.5$ dB and $3.75$ dB for $4$-D $16$-QAM, $6$-D $16$-QAM, and $4$-D $64$-QAM, respectively in  fast Rayleigh fading channel.    The improvement is even more significant compared to random mappings.  \color{black}

\begin{table}[h!]
\caption{Comparison of $\Phi_{fr}(\mu,\mb{\chi})$ at low and high SNR values. }
\centering
\resizebox{0.9\columnwidth}{!}{%
\begin{tabular}{|c||c|c||c|c|}
\hline
\multirow{2}{*}{Mapping}&\multicolumn{2}{c||}{$N = 2$}&\multicolumn{2}{c|}{$N = 3$}\\
\cline{2-5}
  & $\Phi_{fr}(\mu,\mb{\chi})$, $SNR = 2$ dB & $\Phi_{fr}(\mu,\mb{\chi})$, $SNR = 10$ dB & $\Phi_{fr}(\mu,\mb{\chi})$ , $SNR = 2$ dB & $\Phi_{fr}(\mu,\mb{\chi})$ , $SNR = 10$ dB \\
\hline\hline

Random MD $16$-QAM &  4.2146 &   27.807 & 8.5527 & 131.19 \\ \hline

BSA MD $16$-QAM &  9.0667 &   154.63 & 23.699 & 1422.8 \\ \hline

Proposed MD 16-QAM & 10.212 & 209.86 & 35.434 & 3474.2 \\ \thickhline

Random MD $64$-QAM &  5.9020 &   45.960 & 14.267 & 305.68 \\ \hline

BSA MD $64$-QAM &  15.774 & 329.58 & - & - \\ \hline

Proposed MD $64$-QAM & 18.109 & 448.25 & 82.458 & 10483  \\ \hline
\end{tabular}
\label{Parameters_FastFading}
}
\end{table}

\begin{figure}[H]
\centering
\includegraphics[width= 0.6\columnwidth,viewport= 20mm 0mm 320mm 225mm,clip]{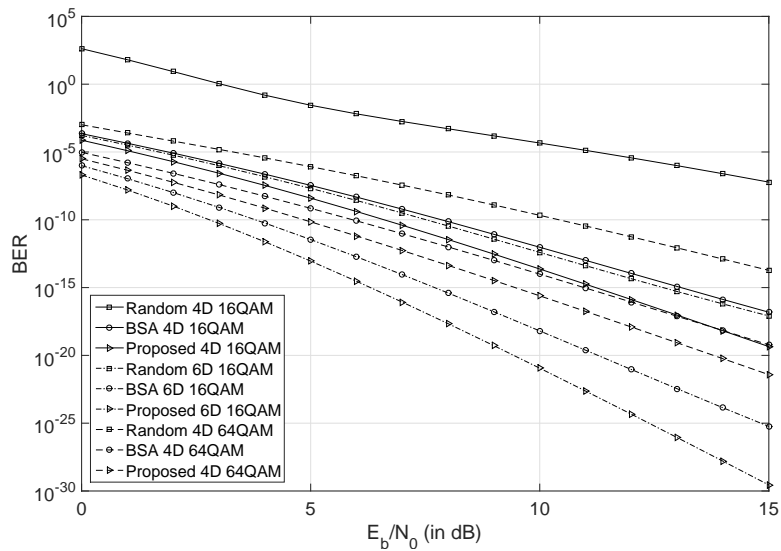}
\caption{Error-floor bounds performance in fast Rayleigh fading channel.}
\label{EF_Fast_Rayleigh}
\end{figure}

\subsection{Convergence Behaviour} 
\begin{figure}[H]
\centering
\includegraphics[width= 0.6\columnwidth,viewport= 20mm 0mm 320mm 225mm,clip]{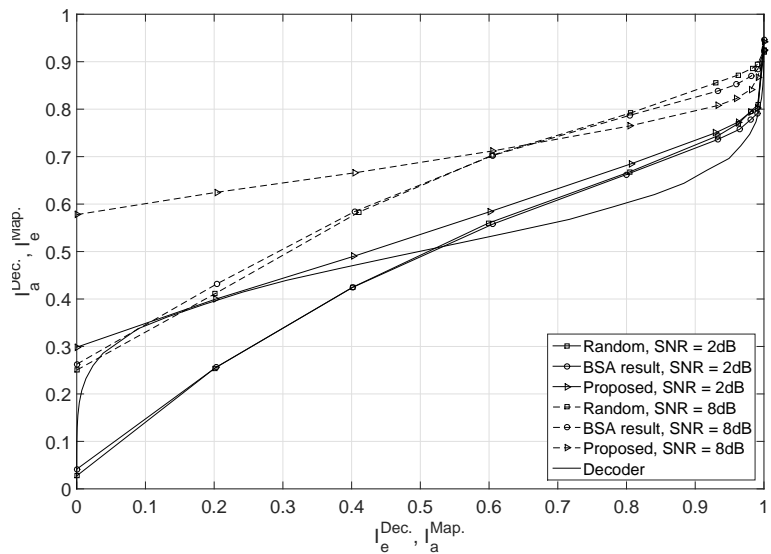}
\caption{ EXIT chart for our  proposed mapping, the mapping found by BSA,  and  random mapping  \color{black}for $4$D $64$-QAM in  AWGN channel.}
\label{EXIT_AWGN}
\end{figure}

Extrinsic information transfer chart (known as EXIT chart) \cite{EXIT} is a suitable technique to investigate the convergence behavior of BICM-ID systems.  The area between the decoder curve and the demapper curve in EXIT chart is referred to as EXIT tunnel \cite{Rob_EXIT}. To improve the BER performance through the iterative process, BICM-ID needs to provide an open Exit tunnel. 
Fig. \ref{EXIT_AWGN} depicts the EXIT charts for BICM-ID using various $4$D $64$-QAM mappings in  AWGN channel.  For brevity, we plot the EXIT chart only for AWGN channels.  The results are very similar for other considered channel models.  
It is obvious from this figure that at SNR equal to $2$dB, BICM-ID with our  proposed mapping exhibits  an open tunnel. It means that at SNR values greater than $2$ dB, iterative decoding with our proposed mapping   improves the BER performance. However, corresponding value of SNR using BSA MD mapping   or random mappings \color{black}  is about $8$ dB. As a result, the BER performance of BICM-ID systems with our proposed mapping starts to  improve through the iterative decoding process  about $6$ dB earlier  than   those of  BSA mapping  and random mapping. \color{black} 
 This also explains the reason of the  early turbo cliff in  BER curves  which can be observed in Fig. \ref{BER_AWGN}. 
 
\section{Conclusion}
\label{conc.}
We have proposed a systematic     method to design  MD mappings for BICM-ID systems using  $16$- and $64$-QAM  constellations. The innovativeness of our proposed mapping  method is that it   can  efficiently generate  MD mappings  using $16$- and $64$-QAM.   Presented numerical results have  shown that in comparison with the well-known BSA  mappings   and random mappings, \color{black}   our resulted mappings outperform  significantly in AWGN,    block-fading \color{black}  and  fast fading channels in the BER  range of practical interest.   Compared to the BSA mappings for a target BER of $10^{-6}$ \color{black}, our mappings can save up to { $3.5$} dB, { $3$} dB, and  $3.75$ dB transmit signal power over  AWGN,   block-fading \color{black} and fast fading channels, respectively.   The corresponding performance gains  are larger compared to  random mappings. \color{black}    The proposed mappings also have  improved  error-floor  performance compared to   random mappings and \color{black} the mappings obtained  by BSA.

\color{black}
\end{document}